\documentclass[12pt,a4paper]{article}
\makeatletter
\def\eqnarray{%
\stepcounter{equation}%
\let\@currentlabel=\theequation
\global\@eqnswtrue
\global\@eqcnt\z@
\tabskip\@centering
\let\\=\@eqncr
$$\halign to \displaywidth\bgroup\@eqnsel\hskip\@centering
$\displaystyle\tabskip\z@{##}$&\global\@eqcnt\@ne
\hfil$\displaystyle{{}##{}}$\hfil
&\global\@eqcnt\tw@$\displaystyle\tabskip\z@{##}$\hfil
\tabskip\@centering&\llap{##}\tabskip\z@\cr}
\makeatother
\newcommand{\kansu}[2]{{{#1}\!\left({#2}\right)}}
\newcommand{\ket}[1]{{\vert{#1}\rangle}}
\newcommand{\bra}[1]{{\langle{#1}\vert}}

\newcommand{\braket}[2]{{\langle{#1}\vert{#2}\rangle}}

\newcommand{\calh}{{\cal H}}
\newcommand{\calm}{{\cal M}}
\newcommand{\cala}{{\cal A}}
\newcommand{\calf}{{\cal F}}

\newcommand{\fukuso}{{\mathbf C}}

\newcommand{\futon}{{\bf N}}

\newcommand{\stm}{{St_m}}
\newcommand{\grm}{{Gr_m}}
\newcommand{\eem}{{E_m}}

\newcommand{\azetta}[1]{{\vert{#1}\vert}}
\newcommand{\szetta}[1]{{\vert{#1}\vert}^{2}}
\newcommand{\qzetta}[1]{{\vert{#1}\vert}^{4}}

\newcommand{\zetta}{{\vert z\vert}}
\newcommand{\wetta}{{\vert w\vert}}

\begin{document}

\title{\sl Geometry of Coherent States : Some Examples 
           of Calculations of Chern--Characters}
\author{
  Kazuyuki FUJII
  \thanks{E-mail address : fujii@math.yokohama-cu.ac.jp}\  
  \thanks{Home-page : http://fujii.sci.yokohama-cu.ac.jp}\\
  Department of Mathematical Sciences\\
  Yokohama City University\\
  Yokohama 236-0027\\ 
  JAPAN
  }
\date{}
\maketitle\thispagestyle{empty}
%
%
%  gaiyou
%
%
\begin{abstract}
  First we make a brief review of coherent states and prove that  
  the resolution of unity can be obtained by the 1--st Chern character
  of some bundle.   
  Next we define a Grassmann manifold for a set of coherent 
  states and construct the pull--back bundle making use of a 
  projector from the parameter space to this Grassmann manifold. 
  We study some geometric properties (Chern--characters mainly) of 
  these bundles. 

  Although the calculations of Chern--characters are in general not easy, 
  we can perform them for the special cases. In this paper we 
  report our calculations and propose some interesting problems 
  to be solved in the near future. 
\end{abstract}

\newpage

%
%
%     Honbun
%
%

\section{Introduction}

Coherent states or generalized coherent states play very important role in 
quantum physics, in particular, quantum optics, see \cite{MW} or 
\cite{KS} and its references.  
They also play an important one in mathematical physics. See the book 
\cite{AP}. For example, they are very useful in performing 
stationary phase approximations to path integral, see \cite{FKSF1}, 
\cite{FKSF2} and \cite{FKS}. 

In this paper we study coherent states from the geometric point of 
view. Namely for a set of coherent states satisfying a certain condition 
we can define a projector from a manifold consisting of parameters to 
(infinite--dimensional) Grassmann manifold $\cdots$ a kind of classifying 
map in K-Theory $\cdots$ and construct a pull--back bundle on the manifold.  
Making use of this we can calculate several geometric quantities, see 
for example \cite{MN}. 

In this paper we mainly focus on Chern--characters because they play an 
very important role in global geometry. But their calculations are not 
so easy. Our calculations are given only for $m=1,\ 2$ (see the section 3). 
Even the case $m=2$ the calculations are complicated enough. 
We leave the case $m=3$ to the readers. 

But it seems to the author that our calculations for $m=2$ suggest some  
deep relation to recent non--commutative differential geometry or 
non--commutative field theory (see the section 4). But this is beyond 
the scope of this paper. We need further study. 

By the way the hidden aim of this paper is to apply the results in this 
paper to Quantum Computation (QC) and Quantum Information Theory (QIT) if 
possible. As for QC or QIT see \cite{LPS}, \cite{AH} and \cite{KF1} for 
general introduction. 
We are in particular interested in Holonomic Quantum Computation, see 
\cite{KF2}--\cite{KF6}. We are also interested in Homodyne Tomography 
\cite{GLM}, \cite{MP} and Quantum Cryptgraphy \cite{KB}, \cite{BW}.

In the forthcoming paper we want to discuss the applications. 

\vspace{10mm}
\section{Coherent States and Grassmann Manifolds}

\subsection{Coherent States}

We make a brief review of some basic properties of coherent operators 
within our necessity, \cite{KS} and \cite{AP}.

Let $a(a^\dagger)$ be the annihilation (creation) operator of the harmonic 
oscillator.
If we set $N\equiv a^\dagger a$ (:\ number operator), then
\begin{equation}
  \label{eq:2-1}
  [N,a^\dagger]=a^\dagger\ ,\
  [N,a]=-a\ ,\
  [a^\dagger, a]=-\mathbf{1}\ .
\end{equation}
Let $\calh$ be a Fock space generated by $a$ and $a^\dagger$, and
$\{\ket{n}\vert\  n\in\futon\cup\{0\}\}$ be its basis.
The actions of $a$ and $a^\dagger$ on $\calh$ are given by
\begin{equation}
  \label{eq:2-2}
  a\ket{n} = \sqrt{n}\ket{n-1}\ ,\
  a^{\dagger}\ket{n} = \sqrt{n+1}\ket{n+1}\ ,
  N\ket{n} = n\ket{n}
\end{equation}
where $\ket{0}$ is a normalized vacuum ($a\ket{0}=0\  {\rm and}\  
\langle{0}\vert{0}\rangle = 1$). From (\ref{eq:2-2})
state $\ket{n}$ for $n \geq 1$ are given by
\begin{equation}
  \label{eq:2-3}
  \ket{n} = \frac{(a^{\dagger})^{n}}{\sqrt{n!}}\ket{0}\ .
\end{equation}
These states satisfy the orthogonality and completeness conditions
\begin{equation}
  \label{eq:2-4}
   \langle{m}\vert{n}\rangle = \delta_{mn}\ ,\quad \sum_{n=0}^{\infty}
   \ket{n}\bra{n} = \mathbf{1}\ . 
\end{equation}

 Let us state coherent states. For the normalized state $\ket{z} \in 
\calh \ {\rm for}\  z \in \fukuso$ the following three conditions are 
equivalent :
\begin{eqnarray}
  \label{eq:2-5-1}
 &&(\mbox{i})\quad a\ket{z} =  z\ket{z}\quad {\rm and}\quad 
      \langle{z}\vert{z}\rangle = 1  \\
  \label{eq:2-5-2}
 &&(\mbox{ii})\quad  \ket{z} =  \mbox{e}^{- \vert{z}\vert^{2}/2} 
          \sum_{n=0}^{\infty}\frac{z^{n}}{\sqrt{n!}}\ket{n} = 
          \mbox{e}^{- \vert{z}\vert^{2}/2}e^{za^{\dagger}}\ket{0} \\
  \label{eq:2-5-3}
 &&(\mbox{iii})\quad  \ket{z} =  \mbox{e}^{za^{\dagger}- \bar{z}a}\ket{0}. 
\end{eqnarray}
In the process from (\ref{eq:2-5-2}) to (\ref{eq:2-5-3}) 
we use the famous elementary Baker-Campbell-Hausdorff formula
\begin{equation}
  \label{eq:2-6}
 \mbox{e}^{A+B}=\mbox{e}^{-\frac1{2}[A,B]}\mbox{e}^{A}\mbox{e}^{B}
\end{equation}
whenever $[A,[A,B]] = [B,[A,B]] = 0$, see \cite{KS} or \cite{AP}.  
This is the key formula.

\noindent{\bfseries Definition}\quad The state $\ket{z}$ that 
satisfies one of (i) or (ii) or (iii) above is called the coherent state.

\noindent
The important feature of coherent states is the following partition 
(resolution) of unity.
\begin{equation}
  \label{eq:2-7}
  \int_{\fukuso} \frac{[d^{2}z]}{\pi} \ket{z}\bra{z} = 
  \sum_{n=0}^{\infty} \ket{n}\bra{n} = \mathbf{1}\ ,
\end{equation}
where we have put $[d^{2}z] = d(\mbox{Re} z)d(\mbox{Im} z)$ for simplicity. 
We note that 
\begin{equation}
   \label{eq:2-a}
  \braket{z}{w} = \mbox{e}^{-\frac{1}{2}\zetta^2 -\frac{1}{2}\wetta^2 +
                             \bar{z}w} \Longrightarrow 
  \vert{\braket{z}{w}}\vert=
  \mbox{e}^{-\frac{1}{2}\vert{z-w}\vert^2},\  
  \braket{w}{z}=\overline{\braket{z}{w}},
\end{equation}
so $\vert{\braket{z}{w}}\vert < 1$ if $z\ne w$ and 
$\vert{\braket{z}{w}}\vert \ll 1$ if $z$ and $w$ are separated enough. 
We will use this fact in the following. 

Since the operator
\begin{equation}
  \label{eq:2-8}
      U(z) = \mbox{e}^{za^{\dagger}- \bar{z}a}
      \quad \mbox{for} \quad z \in \fukuso  
\end{equation}
is unitary, we call this a (unitary) coherent operator. For these 
operators the following properties are crucial.
For  $z,\ w \in \fukuso$
\begin{eqnarray}  
  \label{eq:2-9-1} 
 && U(z)U(w) = \mbox{e}^{z\bar{w}-\bar{z}w}\ U(w)U(z), \\
  \label{eq:2-9-2}
 && U(z+w) = \mbox{e}^{-\frac{1}{2}(z\bar{w}-\bar{z}w)}\ U(z)U(w).
\end{eqnarray}
Here we list some basic properties of this operator.

\vspace{5mm}

\noindent{\bfseries (a) Glauber Formula}\quad  Let $A$ be any observable. 
Then we have 
\begin{equation}
   \label{eq:2-10}
   A = \int_{\fukuso}\frac{[d^{2}z]}{\pi}\mbox{Tr}[AU^{\dagger}(z)]U(z)
\end{equation}
This formula plays an important role in the field of homodyne tomography, 
\cite{GLM} and \cite{MP}.

\vspace{5mm}

\noindent{\bfseries (b) Projection on Coherent State}\quad  
The projection on coherent state $\ket{z}$ is given by $\ket{z}\bra{z}$. 
But this projection has an interesting expression : 
\begin{equation}
   \label{eq:2-11}
          \ket{z}\bra{z} = :\mbox{e}^{-(a-z)^{\dagger}(a-z)}:
\end{equation}
where the notation $:\ :$ means normal ordering. 

\par \noindent
This formula has been used in the field of quantum cryptgraphy, \cite{KB} 
and \cite{BW}. We note that 
\[
          \ket{z}\bra{w} \ne :\mbox{e}^{-(a-z)^{\dagger}(a-w)}:
\]
for $z,\ w \in \fukuso$ with $z\ne w$.

\vspace{10mm}
\subsection{Infinite Grassmann Manifolds and Chern--Characters}

Let $\calh$ be a separable Hilbert space over $\fukuso$.
For $m\in{\bf N}$, we set
\begin{equation}
  \label{eq:stmh}
  \kansu{\stm}{\cal H}
  \equiv
  \left\{
    V=\left(v_1,\cdots,v_m\right)
    \in
    \calh\times\cdots\times\calh\ \vert\  V^\dagger V \in GL(m;\fukuso)
  \right\}\ .
\end{equation}
This is called a (universal) Stiefel manifold.
Note that the general linear group $GL(m)\equiv GL(m;\fukuso)$ acts on 
$\kansu{\stm}{\calh}$ from the right :
\begin{equation}
  \label{eq:stmsha}
  \kansu{\stm}{\calh}\times\kansu{GL}{m}
  \longrightarrow
  \kansu{\stm}{\calh}\  :\  \left( V,a\right)\longmapsto Va\ .
\end{equation}
Next we define a (universal) Grassmann manifold
\begin{equation}
  \kansu{\grm}{\calh}
  \equiv
  \left\{
    X\in\kansu{M}{\calh}\ \vert\ 
    X^2=X, X^\dagger=X\  \mathrm{and}\  \mathrm{tr}X=m\right\}\ ,
\end{equation}
where $M(\calh)$ denotes a space of all bounded linear operators on $\calh$.
Then we have a projection
\begin{equation}
  \label{eq:piteigi}
  \pi : \kansu{\stm}{\calh}\longrightarrow\kansu{\grm}{\calh}\ ,
  \quad \kansu{\pi}{V}\equiv V(V^{\dagger}V)^{-1}V^\dagger\ ,
\end{equation}
compatible with the action (\ref{eq:stmsha}) 
($\kansu{\pi}{Va}=Va(a^{\dagger}V^{\dagger}Va)^{-1}(Va)^\dagger
=\kansu{\pi}{V}$).

Now the set
\begin{equation}
  \label{eq:principal}
  \left\{
    \kansu{GL}{m}, \kansu{\stm}{\calh}, \pi, \kansu{\grm}{\calh}
  \right\}\ ,
\end{equation}
is called a (universal) principal $\kansu{GL}{m}$ bundle, 
see \cite{MN} and \cite{KF1}. \quad We set
\begin{equation}
  \label{eq:emh}
  \kansu{\eem}{\cal H}
  \equiv
  \left\{
    \left(X,v\right)
    \in
    \kansu{\grm}{\calh}\times\calh\ \vert\  Xv=v \right\}\ .
\end{equation}
Then we have also a projection 
\begin{equation}
  \label{eq:piemgrm}
  \pi : \kansu{\eem}{\calh}\longrightarrow\kansu{\grm}{\calh}\ ,
  \quad \kansu{\pi}{\left(X,v\right)}\equiv X\ .
\end{equation}
The set
\begin{equation}
  \label{eq:universal}
  \left\{
    \fukuso^m, \kansu{\eem}{\calh}, \pi, \kansu{\grm}{\calh}
  \right\}\ ,
\end{equation}
is called a (universal) $m$--th vector bundle. This vector bundle is 
one associated with the principal $\kansu{GL}{m}$ bundle (\ref{eq:principal}). 

Next let ${\calm}$ be a finite or infinite dimensional differentiable manifold 
and the map 
\begin{equation}
  \label{eq:projector}
   P : {\calm} \longrightarrow \kansu{\grm}{\calh}
\end{equation} 
be given (called a projector). Using this $P$ we can make the bundles 
(\ref{eq:principal}) and (\ref{eq:universal}) pullback over ${\calm}$ :
\begin{eqnarray}
  \label{eq:hikimodoshi1}
  &&\left\{\kansu{GL}{m},\widetilde{St}, \pi_{\widetilde{St}}, {\calm}\right\}
  \equiv
  P^*\left\{\kansu{GL}{m}, \kansu{\stm}{\calh}, \pi, 
  \kansu{\grm}{\calh}\right\}
  \ , \\
  \label{eq:hikimodoshi2}
  &&\left\{\fukuso^m,\widetilde{E}, \pi_{\widetilde{E}}, {\calm}\right\}
  \equiv
  P^*\left\{\fukuso^m, \kansu{\eem}{\calh}, \pi, \kansu{\grm}{\calh}\right\}
  \ , 
\end{eqnarray}

\[    
   \matrix{
    \kansu{GL}{m}&&\kansu{GL}{m}\cr
    \Big\downarrow&&\Big\downarrow\cr
    \widetilde{St}&\longrightarrow&\kansu{\stm}{\calh}\cr
    \Big\downarrow&&\Big\downarrow\cr
    {\calm}&\stackrel{P}{\longrightarrow}&\kansu{\grm}{\calh}\cr
           } \qquad \qquad  
   \matrix{
    \fukuso^m&&\fukuso^m\cr
    \Big\downarrow&&\Big\downarrow\cr
    \widetilde{E}&\longrightarrow&\kansu{E_m}{\calh}\cr
    \Big\downarrow&&\Big\downarrow\cr
    {\calm}&\stackrel{P}{\longrightarrow}&\kansu{\grm}{\calh}\cr
           } 
\]
\par \vspace{5mm} \noindent
see \cite{MN}. (\ref{eq:hikimodoshi2}) is of course a vector bundle 
associated with (\ref{eq:hikimodoshi1}).

For this bundle the (global) curvature ($2$--) form $\bf\Omega$ is given by 
\begin{equation}
  \label{eq:curvature}
  {\bf\Omega}=PdP\wedge dP 
\end{equation}
making use of (\ref{eq:projector}), where $d$ is the differential form 
on $\bf\Omega$. 
For the bundles Chern--characters play an essential role in several geometric 
properties. In this case Chern--characters are defined (see \cite{MN}, 
Section 11) by making use of 
\begin{equation}
  \label{eq:Chern--classes}
  {\bf\Omega},\ {\bf\Omega}^2,\ \cdots,\ {\bf\Omega}^{m/2},
\end{equation}
where we have assumed that $m=\mbox{dim}{\calm}$ is even. 
We note that ${\bf\Omega}^2={\bf\Omega}\wedge {\bf\Omega}$, etc.

\par \noindent  
In this paper we don't take the trace of (\ref{eq:Chern--classes}) 
, so it may be better to call them pre--Chern characters.  
We want to caluculate them directly. 

To calculate these quantities in infinite--dimensional cases is not so 
easy. In the next section let us calculate them in the special cases.

We now define our projectors for the latter aim. For $z_1, z_2, \cdots, 
z_m \in \fukuso$ we set 
\begin{equation}
  \label{eq:a set of coherent states}   
   V_{m}({\bf z})=(\ket{z_1},\ket{z_2}, \cdots, \ket{z_m})\equiv V_{m}
\end{equation}
where ${\bf z}=(z_1, z_2, \cdots, z_m)$. 
Since ${V_{m}}^{\dagger}V_{m}=(\braket{z_i}{z_j}) \in M(m,\fukuso)$,  
we define 
\begin{equation}
  \label{eq:domain}
    {\cal D}_m \equiv \{{\bf z}\in \fukuso^{m}\ \vert\ 
               \mbox{det}({V_{m}}^{\dagger}V_{m}) \ne 0 \}
    = \{{\bf z}\in \fukuso^{m}\ \vert\ {V_{m}}^{\dagger}V_{m}\in GL(m)\}. 
\end{equation}
For example ${V_{1}}^{\dagger}V_{1}=1$ for $m=1$, and for $m=2$ 
\[
   \mbox{det}({V_{2}}^{\dagger}V_{2}) = 
  \left|
    \begin{array}{cc}
        1 & a \\
        \bar{a} & 1 
    \end{array}
  \right|
   = 1-\vert{a}\vert^{2} \geq 0
\]
where $a=\braket{z_1}{z_2}$. So from (\ref{eq:2-a}) we have 
\begin{eqnarray}
  \label{eq:condition-1}   
   {\cal D}_1 &=&\{z \in \fukuso \ \vert\ \mbox{no conditions}\} 
               = \fukuso ,  \\
  \label{eq:condition-2}   
   {\cal D}_2 &=&\{(z_1,z_2) \in \fukuso^2 \ \vert\ z_1\ne z_2 \}.
\end{eqnarray}
For ${\cal D}_m\ (m\geq 3)$ it is not easy for us to give a simple 
condition like (\ref{eq:condition-2}). 
\begin{flushleft}
{\bf Problem}\quad For the case $m=3$ make the condition (\ref{eq:domain}) 
clear like (\ref{eq:condition-2}). 
\end{flushleft}
At any rate 
$
V_{m} \in \kansu{\stm}{\calh} \ \mbox{for}\ {\bf z} \in {\cal D}_m 
$ .
Now let us define our projector $P$ as follows : 
\begin{eqnarray}
  \label{eq:real-projector}
  P : {\cal D}_m \longrightarrow \kansu{\grm}{\calh}\ , \quad 
    P({\bf z})= V_{m}(V_{m}^{\dagger}V_{m})^{-1}V_{m}^{\dagger}\ .
\end{eqnarray}
In the following we set $V=V_{m}$ for simplicity.  Let us calculate 
(\ref{eq:curvature}). Since 
\begin{eqnarray}
   dP&=&dV(V^{\dagger}V)^{-1}V^{\dagger}- V(V^{\dagger}V)^{-1}
      (dV^{\dagger}V+V^{\dagger}dV)(V^{\dagger}V)^{-1}V^{\dagger}+ 
      V(V^{\dagger}V)^{-1}dV^{\dagger} \nonumber \\
    &=&V(V^{\dagger}V)^{-1}dV^{\dagger}
     \{
        {\bf 1}-V(V^{\dagger}V)^{-1}V^{\dagger}
     \}
     + \{{\bf 1}-V(V^{\dagger}V)^{-1}V^{\dagger}\}dV(V^{\dagger}V)^{-1}
      V^{\dagger} \nonumber 
\end{eqnarray}
where $d=\sum_{j=1}^{m}\left(dz_{j}\frac{\partial}{\partial z_{j}}+ 
d{\bar z_{j}}\frac{\partial}{\partial {\bar z_{j}}}\right)$, we have 
\[
  PdP=V(V^{\dagger}V)^{-1}dV^{\dagger}
      \{{\bf 1}-V(V^{\dagger}V)^{-1}V^{\dagger}\}.
\]
after some calculation. Therefore we obtain  
\begin{equation}
  \label{eq:curvature-local}
  PdP\wedge dP=V(V^{\dagger}V)^{-1}[dV^{\dagger}\{
               {\bf 1}-V(V^{\dagger}V)^{-1}V^{\dagger}
               \}dV](V^{\dagger}V)^{-1}V^{\dagger}\ .
\end{equation}
Our main calculation is $dV^{\dagger}
\{{\bf 1}-V(V^{\dagger}V)^{-1}V^{\dagger}\}dV$, which is rewritten as 
\begin{equation}
  \label{eq:curvature-decomposition}
  dV^{\dagger}\{{\bf 1}-V(V^{\dagger}V)^{-1}V^{\dagger}\}dV=
  {[\{{\bf 1}-V(V^{\dagger}V)^{-1}V^{\dagger}\}dV]}^{\dagger}\ 
  [\{{\bf 1}-V(V^{\dagger}V)^{-1}V^{\dagger}\}dV] 
\end{equation}
since $Q \equiv {\bf 1}-V(V^{\dagger}V)^{-1}V^{\dagger}$ is also a projector
($Q^2=Q$ and $Q^{\dagger}=Q$). Therefore the first step for us is to 
calculate the term 
\begin{equation}
  \label{eq:main-term}
  \{{\bf 1}-V(V^{\dagger}V)^{-1}V^{\dagger}\}dV\ .
\end{equation}
\par \noindent 
Let us summarize our process of calculations : 
\begin{eqnarray}
  &&\mbox{1--st step}\qquad  \{{\bf 1}-V(V^{\dagger}V)^{-1}V^{\dagger}\}dV\ ,
      \nonumber \\
  &&\mbox{2--nd step}\qquad 
    dV^{\dagger}\{{\bf 1}-V(V^{\dagger}V)^{-1}V^{\dagger}\}dV\  \cdots 
    (\ref{eq:curvature-decomposition}),  \nonumber \\
  &&\mbox{3--rd step}\qquad 
    V(V^{\dagger}V)^{-1}
    [dV^{\dagger}\{{\bf 1}-V(V^{\dagger}V)^{-1}V^{\dagger}\}dV]   
    (V^{\dagger}V)^{-1}V^{\dagger}\  \cdots 
    (\ref{eq:curvature-local}).  \nonumber 
\end{eqnarray}

\vspace{1cm}
\section{Examples of Calculations of Chern--Characters}

In this section we calculate the Chern--characters only for the cases $m=1$
and $m=2$. Even for $m=2$ the calculation is complicated enough. 
For $m\geq 3$ calculations become miserable.

\subsection{M=1}

In this case $\braket{z}{z}=1$, so our projector is very simple to be 
\begin{equation}
  \label{eq:1-projector} 
     P(z)=\ket{z}\bra{z}. 
\end{equation}
In this case the calculation of curvature is relatively simple. 
From (\ref{eq:curvature-local}) we have 
\begin{equation}
  \label{eq:1-curvature} 
  PdP\wedge dP=\ket{z}\{d\bra{z}({\bf 1}-\ket{z}\bra{z})d\ket{z}\}\bra{z}
         =\ket{z}\bra{z}\{d\bra{z}({\bf 1}-\ket{z}\bra{z})d\ket{z}\}.
\end{equation}
Since $\ket{z}=\mbox{exp}(-\frac{1}{2}\zetta^2)\mbox{exp}(za^{\dagger})
\ket{0}$ by (\ref{eq:2-5-2}), 
\[
 d\ket{z}=\left\{
                 (a^{\dagger}-\frac{{\bar z}}{2})dz-\frac{z}{2}d{\bar z}
          \right\}\ket{z}
         =\left\{
                 a^{\dagger}dz-\frac{1}{2}({\bar z}dz+zd{\bar z})
          \right\}\ket{z}
         =\left\{
                 a^{\dagger}dz-\frac{1}{2}d(\zetta^2)
          \right\}\ket{z},
\]
so that 
\[
  ({\bf 1}-\ket{z}\bra{z})d\ket{z}=
     ({\bf 1}-\ket{z}\bra{z})a^{\dagger}\ket{z}dz =
     (a^{\dagger}-\bra{z}a^{\dagger}\ket{z})\ket{z}dz =
     (a-z)^{\dagger}dz\ket{z}
\]
because $({\bf 1}-\ket{z}\bra{z})\ket{z}={\bf 0}$. Similarly 
$d\bra{z}({\bf 1}-\ket{z}\bra{z})=\bra{z}(a-z) d{\bar z}$. 
\par \noindent 
Let us summarize : 
\begin{equation}
  \label{eq:ralations}
   ({\bf 1}-\ket{z}\bra{z})d\ket{z}=(a-z)^{\dagger}dz\ket{z}, \quad 
   d\bra{z}({\bf 1}-\ket{z}\bra{z})=\bra{z}(a-z) d{\bar z}\ .
\end{equation}
Now we are in a position to determine the curvature form 
(\ref{eq:1-curvature}).
\[
  d\bra{z}({\bf 1}-\ket{z}\bra{z})d\ket{z}=
  \bra{z}(a-z)(a-z)^{\dagger}\ket{z}d{\bar z}\wedge dz=
  d{\bar z}\wedge dz
\]
after some algebra. Therefore
\begin{equation}
  \label{eq:result-1}
 {\bf\Omega}= PdP\wedge dP=\ket{z}\bra{z}d{\bar z}\wedge dz\ .
\end{equation}
From this result we know 
\[
  \frac{{\bf\Omega}}{2\pi i}=\ket{z}\bra{z}\frac{dx\wedge dy}{\pi}
\]
when $z=x+iy$. This just gives the resolution of unity in (\ref{eq:2-7}).

\vspace{10mm}
\subsection{M=2 $\cdots$ Main Result}

First of all let us determine the projector. Since $V=(\ket{z_1},\ket{z_2})$ 
we have easily 
\begin{eqnarray}
   P(z_1, z_2)&=&\left(\ket{z_1},\ket{z_2}\right)
 {\left(
    \begin{array}{cc}
        1 & \braket{z_1}{z_2} \\
        \braket{z_2}{z_1} & 1 
    \end{array}
  \right)}^{-1}
 {\left(
    \begin{array}{c}
       \bra{z_1} \\
       \bra{z_2}  
    \end{array}
  \right)}   \nonumber \\
 &=&\frac{1}{1-\vert{\braket{z_1}{z_2}}\vert^{2}}
   (\ket{z_1},\ket{z_2})
 {\left(
    \begin{array}{cc}
        1 & -\braket{z_1}{z_2} \\
        -\braket{z_2}{z_1} & 1 
    \end{array}
  \right)}
 {\left(
    \begin{array}{c}
       \bra{z_1} \\
       \bra{z_2}  
    \end{array}
  \right)}   \nonumber \\
 &=&\frac{1}{1-\vert{\braket{z_1}{z_2}}\vert^{2}}
   \left(\ket{z_1}\bra{z_1}-\braket{z_2}{z_1}\ket{z_2}\bra{z_1}
         -\braket{z_1}{z_2}\ket{z_1}\bra{z_2}+\ket{z_2}\bra{z_2}
   \right). \quad
\end{eqnarray}
\par \noindent
Let us calculate (\ref{eq:main-term}) : Since 
\[
  dV=(d\ket{z_1},d\ket{z_2})=
\left(
  \left\{a^{\dagger}dz_1-\frac{1}{2}d(\szetta{z_1})\right\}\ket{z_1}, 
  \left\{a^{\dagger}dz_2-\frac{1}{2}d(\szetta{z_2})\right\}\ket{z_2}
\right)
\]
the straightforward calculation leads 
\begin{equation}
  \{{\bf 1}-V(V^{\dagger}V)^{-1}V^{\dagger}\}dV=(K_1, K_2) 
\end{equation}
where 
\begin{eqnarray}
  K_1&=&\left\{(a-z_2)^{\dagger}-
        \frac{{\bar z_1}-{\bar z_2}}{1-\vert{\braket{z_1}{z_2}}\vert^{2}}
        \right\}dz_1\ket{z_1}
     +\frac{{\braket{z_2}{z_1}}({\bar z_1}-{\bar z_2})}
           {{1-\vert{\braket{z_1}{z_2}}\vert^{2}}}
        dz_1\ket{z_2}, 
    \nonumber \\
 &&{}  \nonumber \\
  K_2&=&\frac{{\braket{z_1}{z_2}}({\bar z_2}-{\bar z_1})}
             {{1-\vert{\braket{z_1}{z_2}}\vert^{2}}}dz_2\ket{z_1}
       +\left\{(a-z_1)^{\dagger}-
        \frac{{\bar z_2}-{\bar z_1}}{1-\vert{\braket{z_1}{z_2}}\vert^{2}}
        \right\}dz_2\ket{z_2}.
\end{eqnarray}
\par \vspace{5mm} \noindent
Therefore by (\ref{eq:curvature-decomposition})
\begin{equation}
  \label{eq:matrix-form}
  dV^{\dagger}\{{\bf 1}-V(V^{\dagger}V)^{-1}V^{\dagger}\}dV=
  \left(
    \begin{array}{cc}
       F_{11}& F_{12} \\
       F_{21}& F_{22} 
    \end{array}
  \right)  
\end{equation}
where 
\begin{eqnarray}
  F_{11}&=&\left\{1-\frac{\szetta{z_1-z_2}\szetta{\braket{z_1}{z_2}}}
                    {1-\szetta{\braket{z_1}{z_2}}}
           \right\}d{\bar z_1}\wedge dz_1\ , \ 
  F_{12}=  \braket{z_1}{z_2}
           \left\{1-\frac{\szetta{z_1-z_2}}
                    {1-\szetta{\braket{z_1}{z_2}}}
           \right\}d{\bar z_1}\wedge dz_2\ ,   \nonumber \\
&&{} \nonumber \\
  F_{21}&=&\braket{z_2}{z_1}
           \left\{1-\frac{\szetta{z_1-z_2}}
                    {1-\szetta{\braket{z_1}{z_2}}}
           \right\}d{\bar z_2}\wedge dz_1\ , \ 
  F_{22}=  \left\{1-\frac{\szetta{z_1-z_2}\szetta{\braket{z_1}{z_2}}}
                    {1-\szetta{\braket{z_1}{z_2}}}
           \right\}d{\bar z_2}\wedge dz_2\ .  \nonumber \\
        &&{}
\end{eqnarray}
\par \noindent 
Now we are in a position to determine the curvature form 
(\ref{eq:curvature-local}). 
Since 
\begin{eqnarray}
  V(V^{\dagger}V)^{-1}&=&
   \frac{1}{1-\vert{\braket{z_1}{z_2}}\vert^{2}}
   (\ket{z_1}-\braket{z_2}{z_1}\ket{z_2}, 
      \ket{z_2}-\braket{z_1}{z_2}\ket{z_1}), \nonumber \\
 (V^{\dagger}V)^{-1}V^{\dagger}&=&
   \frac{1}{1-\vert{\braket{z_1}{z_2}}\vert^{2}}
  {\left(
    \begin{array}{c}
       \bra{z_1}- \bra{z_2}\braket{z_1}{z_2}\\
       \bra{z_2}- \bra{z_1}\braket{z_2}{z_1} 
    \end{array}
  \right)}  \nonumber   
\end{eqnarray}
we obtain 
\begin{eqnarray}
  \label{eq:exact-calculation-2form} 
 &{\bf\Omega}&=PdP\wedge dP   \nonumber \\
 &=&\frac{1}{(1-\vert{\braket{z_1}{z_2}}\vert^{2})^{2}}
   \left(
           \ket{z_1}\bra{z_1}L_1
          -\braket{z_2}{z_1}\ket{z_2}\bra{z_1}L_2
          -\braket{z_1}{z_2}\ket{z_1}\bra{z_2}L_3
          +\ket{z_2}\bra{z_2}L_4
   \right), \quad {}  
\end{eqnarray}
where 
\begin{eqnarray}
  L_1 &=& \left\{1-\frac{\szetta{z_1-z_2}\szetta{\braket{z_1}{z_2}}}
                  {1-\szetta{\braket{z_1}{z_2}}}
          \right\}d{\bar z_1}\wedge dz_1
        - \szetta{\braket{z_1}{z_2}}
            \left\{1-\frac{\szetta{z_1-z_2}}{1-\szetta{\braket{z_1}{z_2}}}
            \right\}d{\bar z_2}\wedge dz_1  \nonumber \\
      &-& \szetta{\braket{z_1}{z_2}}
            \left\{1-\frac{\szetta{z_1-z_2}}{1-\szetta{\braket{z_1}{z_2}}}
            \right\}d{\bar z_1}\wedge dz_2
        + \szetta{\braket{z_1}{z_2}}
         \left\{1-\frac{\szetta{z_1-z_2}\szetta{\braket{z_1}{z_2}}}
                  {1-\szetta{\braket{z_1}{z_2}}}
          \right\}d{\bar z_2}\wedge dz_2   \nonumber \\
      &&{} \nonumber \\
  L_2 &=& \left\{1-\frac{\szetta{z_1-z_2}\szetta{\braket{z_1}{z_2}}}
                  {1-\szetta{\braket{z_1}{z_2}}}
          \right\}d{\bar z_1}\wedge dz_1
         - \left\{1-\frac{\szetta{z_1-z_2}}{1-\szetta{\braket{z_1}{z_2}}}
            \right\}d{\bar z_2}\wedge dz_1  \nonumber \\
      &-&\szetta{\braket{z_1}{z_2}}  
         \left\{1-\frac{\szetta{z_1-z_2}}{1-\szetta{\braket{z_1}{z_2}}}
            \right\}d{\bar z_1}\wedge dz_2
        + \left\{1-\frac{\szetta{z_1-z_2}\szetta{\braket{z_1}{z_2}}}
                  {1-\szetta{\braket{z_1}{z_2}}}
           \right\}d{\bar z_2}\wedge dz_2 \ , \nonumber \\
      &&{} \nonumber \\
  L_3 &=& \left\{1-\frac{\szetta{z_1-z_2}\szetta{\braket{z_1}{z_2}}}
                  {1-\szetta{\braket{z_1}{z_2}}}
          \right\}d{\bar z_1}\wedge dz_1
        - \szetta{\braket{z_1}{z_2}}
            \left\{1-\frac{\szetta{z_1-z_2}}{1-\szetta{\braket{z_1}{z_2}}}
            \right\}d{\bar z_2}\wedge dz_1  \nonumber \\
      &-&   \left\{1-\frac{\szetta{z_1-z_2}}{1-\szetta{\braket{z_1}{z_2}}}
            \right\}d{\bar z_1}\wedge dz_2
        + \left\{1-\frac{\szetta{z_1-z_2}\szetta{\braket{z_1}{z_2}}}
                  {1-\szetta{\braket{z_1}{z_2}}}
           \right\}d{\bar z_2}\wedge dz_2 \ , \nonumber \\
      &&{} \nonumber \\
  L_4 &=& \szetta{\braket{z_1}{z_2}}
         \left\{1-\frac{\szetta{z_1-z_2}\szetta{\braket{z_1}{z_2}}}
                  {1-\szetta{\braket{z_1}{z_2}}}
          \right\}d{\bar z_1}\wedge dz_1
        - \szetta{\braket{z_1}{z_2}}
            \left\{1-\frac{\szetta{z_1-z_2}}{1-\szetta{\braket{z_1}{z_2}}}
            \right\}d{\bar z_2}\wedge dz_1  \nonumber \\
      &-& \szetta{\braket{z_1}{z_2}}
            \left\{1-\frac{\szetta{z_1-z_2}}{1-\szetta{\braket{z_1}{z_2}}}
            \right\}d{\bar z_1}\wedge dz_2 
        + \left\{1-\frac{\szetta{z_1-z_2}\szetta{\braket{z_1}{z_2}}}
                  {1-\szetta{\braket{z_1}{z_2}}}
          \right\}d{\bar z_2}\wedge dz_2 \ . 
\end{eqnarray}
\par \vspace{15mm} \noindent 
This is our main result. Next let us calculate ${\bf\Omega}^2$ (${\bf\Omega}^k
={\bf 0}$ for $k\geq 3$) : From (\ref{eq:exact-calculation-2form}) we obtain 
after long calculation 
\vspace{5mm} 
\begin{eqnarray}
 \label{eq:exact-calculation-4form}
 {\bf\Omega}^2 &=&
 \frac{1}{(1-\vert{\braket{z_1}{z_2}}\vert^{2})^{4}}  
 \left(
           \ket{z_1}\bra{z_1}M_1
          -\braket{z_2}{z_1}\ket{z_2}\bra{z_1}M_2 
          -\braket{z_1}{z_2}\ket{z_1}\bra{z_2}M_3
          +\ket{z_2}\bra{z_2}M_4
 \right) \nonumber \\
 &\times&  dz_1\wedge d{\bar z_1}\wedge dz_2\wedge d{\bar z_2}\ ,  
\end{eqnarray}
where 
\begin{eqnarray}
 M_1&=& \szetta{\braket{z_1}{z_2}}(1-\szetta{\braket{z_1}{z_2}})^2  
       + 2\szetta{\braket{z_1}{z_2}}(1-\szetta{\braket{z_1}{z_2}})
            \szetta{z_1-z_2} \nonumber \\
     &-&\szetta{\braket{z_1}{z_2}}(1+2\szetta{\braket{z_1}{z_2}})
            \qzetta{z_1-z_2}\ , \nonumber \\
&&{}  \nonumber \\
 M_2&=& (1-\szetta{\braket{z_1}{z_2}})^2  
       + 2\szetta{\braket{z_1}{z_2}}(1-\szetta{\braket{z_1}{z_2}})
            \szetta{z_1-z_2} \nonumber \\
     &-&\szetta{\braket{z_1}{z_2}}(2+\szetta{\braket{z_1}{z_2}})
            \qzetta{z_1-z_2}\ , \nonumber \\
&&{}  \nonumber \\
 M_3&=& (1-\szetta{\braket{z_1}{z_2}})^2  
       + 2\szetta{\braket{z_1}{z_2}}(1-\szetta{\braket{z_1}{z_2}})
            \szetta{z_1-z_2} \nonumber \\
     &-&\szetta{\braket{z_1}{z_2}}(2+\szetta{\braket{z_1}{z_2}})
            \qzetta{z_1-z_2}\ , \nonumber \\
&&{}  \nonumber \\
 M_4&=& \szetta{\braket{z_1}{z_2}}(1-\szetta{\braket{z_1}{z_2}})^2  
       + 2\szetta{\braket{z_1}{z_2}}(1-\szetta{\braket{z_1}{z_2}})
            \szetta{z_1-z_2} \nonumber \\
     &-&\szetta{\braket{z_1}{z_2}}(1+2\szetta{\braket{z_1}{z_2}})
            \qzetta{z_1-z_2}\ . 
\end{eqnarray}
This is a second main result in this paper.

\par \noindent
We have calculated the Chern--characters for $m=2$. Since our results are 
in a certain sense ``raw'' (remember that we don't take the trace), one 
can freely ``cook'' them. We leave it to the readers.

\vspace{1cm}
\subsection{Problems}
 
Before concluding this section let us propose problems : 
\begin{flushleft}
{\bf Problem\ 3}\quad For the case of $m=3$ perform the similar calculations ! 
\end{flushleft}
We want to calculate them up to this case. Therefore let us give an explicit 
form to the projector : 
\begin{eqnarray}
   &&P(z_1, z_2, z_3)=\left(\ket{z_1},\ket{z_2},\ket{z_3}\right)
 {\left(
    \begin{array}{ccc}
        1 & \braket{z_1}{z_2} & \braket{z_1}{z_3} \\
        \braket{z_2}{z_1} & 1 & \braket{z_2}{z_3} \\
        \braket{z_3}{z_1} & \braket{z_3}{z_2} & 1 
    \end{array}
  \right)}^{-1}
 {\left(
    \begin{array}{c}
       \bra{z_1} \\
       \bra{z_2} \\
       \bra{z_3}  
    \end{array}
  \right)}   \nonumber \\
    &&\quad =\frac{1}{\mbox{det}M}
     {\bf [}\ 
      \{1-\szetta{\braket{z_2}{z_3}}\}\ket{z_1}\bra{z_1}
     -\{\braket{z_1}{z_2}-\braket{z_1}{z_3}\braket{z_3}{z_2}\}
        \ket{z_1}\bra{z_2}  \nonumber \\
    &&\quad\ -\{\braket{z_1}{z_3}-\braket{z_1}{z_2}\braket{z_2}{z_3}\}
        \ket{z_1}\bra{z_3}
      -\{\braket{z_2}{z_1}-\braket{z_2}{z_3}\braket{z_3}{z_1}\}
        \ket{z_2}\bra{z_1}  \nonumber \\
    &&\quad\  + \{1-\szetta{\braket{z_1}{z_3}}\}\ket{z_2}\bra{z_2}
       -\{\braket{z_2}{z_3}-\braket{z_2}{z_1}\braket{z_1}{z_3}\}
         \ket{z_2}\bra{z_3}  \nonumber \\ 
    &&\quad\  -\{\braket{z_3}{z_1}-\braket{z_3}{z_2}\braket{z_2}{z_1}\}
        \ket{z_3}\bra{z_1}
       -\{\braket{z_3}{z_2}-\braket{z_3}{z_1}\braket{z_1}{z_2}\}
          \ket{z_3}\bra{z_2}  \nonumber \\ 
    &&\quad\  + \{1-\szetta{\braket{z_1}{z_2}}\}\ket{z_3}\bra{z_3}\  
     \ {\bf ]},
\end{eqnarray}
where
\[
  \mbox{det}M=1-\szetta{\braket{z_1}{z_2}}-\szetta{\braket{z_2}{z_3}}
      -\szetta{\braket{z_1}{z_3}}
      +\braket{z_1}{z_2}\braket{z_2}{z_3}\braket{z_3}{z_1}
      +\braket{z_1}{z_3}\braket{z_3}{z_2}\braket{z_2}{z_1}.
\]
Perform the calculations of ${\bf\Omega}$, ${\bf\Omega}^{2}$ and 
${\bf\Omega}^{3}$. \quad Moreover 
\begin{flushleft}
{\bf Problem\ {$\infty$}}\quad For the general case perform the similar 
calculations (if possible).
\end{flushleft}
It seems to the author that the calculations in the general case are 
very hard.

\vspace{10mm}
\section{A Supplement}

In this section we make a brief supplement to the discussion in 
the preceeding section. 

For $V=V_{m}$ in (\ref{eq:a set of coherent states}) we set 
\begin{equation}
{\widetilde V}=V(V^{\dagger}V)^{-1/2}\ ,
\end{equation}
then ${\widetilde V}$ satisfies the equation 
${\widetilde V}^{\dagger}{\widetilde V}=1_{m}$. Then the canonical connection 
form $\cala$ is given by 
\begin{eqnarray}
 \label{eq:connection-form}
  \cala&\equiv&{\widetilde V}^{\dagger}d{\widetilde V}
       =(V^{\dagger}V)^{-1/2}V^{\dagger}dV(V^{\dagger}V)^{-1/2}+
        (V^{\dagger}V)^{1/2}d(V^{\dagger}V)^{-1/2} \nonumber \\
       &=&(V^{\dagger}V)^{1/2}\{(V^{\dagger}V)^{-1}V^{\dagger}dV\}
        (V^{\dagger}V)^{-1/2}+(V^{\dagger}V)^{1/2}d(V^{\dagger}V)^{-1/2}\ , 
\end{eqnarray}
so the (local) curvature form $\calf = d\cala+\cala\wedge \cala$ becomes 
\begin{equation}
 \label{eq:local-curvature-form}
\calf = (V^{\dagger}V)^{-1/2}[dV^{\dagger}\{{\bf 1}-V(V^{\dagger}V)^{-1}
         V^{\dagger}\}dV](V^{\dagger}V)^{-1/2}\ 
\end{equation}
because of 
\begin{eqnarray}
  &&d\{(V^{\dagger}V)^{-1}V^{\dagger}dV\}+ 
   \{(V^{\dagger}V)^{-1}V^{\dagger}dV\}\wedge 
   \{(V^{\dagger}V)^{-1}V^{\dagger}dV\}        \nonumber \\
  &&=(V^{\dagger}V)^{-1}dV^{\dagger}
      \{{\bf 1}-V(V^{\dagger}V)^{-1}V^{\dagger}\}dV\ .  \nonumber 
\end{eqnarray}
That is, $(V^{\dagger}V)^{-1}V^{\dagger}dV$ is a main term of $\cala$. 
By the way the relation between (\ref{eq:curvature-local}) and $\calf$ 
above is given by 
\begin{equation}
    PdP\wedge dP={\widetilde V}\calf {\widetilde V}^{\dagger} \ . 
\end{equation}
For the case of $m=2$ let us calculate the connection form 
(\ref{eq:connection-form}). But since the calculation is very complicated, 
we only calculate the main term in (\ref{eq:connection-form}), which is 
essential to calculate the curvature form as shown above. 
Since 
\begin{equation}
  V^{\dagger}dV=
 {\left(
    \begin{array}{cc}
        {\bar z_1}dz_1-\frac{1}{2}d(\szetta{z_1})& \braket{z_1}{z_2}
        \{{\bar z_1}dz_2-\frac{1}{2}d(\szetta{z_2})\}  \\
       \braket{z_2}{z_1}\{{\bar z_2}dz_1-\frac{1}{2}d(\szetta{z_1})\}&
        {\bar z_2}dz_2-\frac{1}{2}d(\szetta{z_2})
    \end{array}
  \right)}
\end{equation}
we have 
\begin{equation}
 (V^{\dagger}V)^{-1}V^{\dagger}dV=
  \frac{1}{1-\szetta{\braket{z_1}{z_2}}}
 {\left(
    \begin{array}{cc}
      N_{11}& N_{12} \\
      N_{21}& N_{22} 
   \end{array}
  \right)}
\end{equation}
where
\begin{eqnarray}
  N_{11}&=&\left({\bar z_1}-\szetta{\braket{z_1}{z_2}}{\bar z_2}\right)dz_1 - 
      \frac{1}{2}\left(1-\szetta{\braket{z_1}{z_2}}\right)d(\szetta{z_1}),\ 
  N_{12}=\braket{z_1}{z_2}\left({\bar z_1}-{\bar z_2}\right)dz_2,
       \nonumber \\
  N_{21}&=&\braket{z_2}{z_1}\left({\bar z_2}-{\bar z_1}\right)dz_1, \  
  N_{22}=\left({\bar z_2}-\szetta{\braket{z_1}{z_2}}{\bar z_1}\right)dz_2 - 
      \frac{1}{2}\left(1-\szetta{\braket{z_1}{z_2}}\right)d(\szetta{z_2}).
       \nonumber \\ 
  &&{}  
\end{eqnarray}

\par \noindent 
Here we note that
\begin{eqnarray}
\label{eq:v-dagger-v}
  (V^{\dagger}V)^{1/2}&=&
 {\left(
    \begin{array}{cc}
      s& \frac{\braket{z_1}{z_2}}{\azetta{\braket{z_1}{z_2}}}t \\
      \frac{\braket{z_2}{z_1}}{\azetta{\braket{z_1}{z_2}}}t &s
   \end{array}
  \right)},    \\
\label{eq:v-dagger-v-inverse}
  (V^{\dagger}V)^{-1/2}&=&\frac{1}{\sqrt{1-\szetta{\braket{z_1}{z_2}}}}
 {\left(
    \begin{array}{cc}
      s& -\frac{\braket{z_1}{z_2}}{\azetta{\braket{z_1}{z_2}}}t \\
      -\frac{\braket{z_2}{z_1}}{\azetta{\braket{z_1}{z_2}}}t &s
   \end{array}
  \right)}, 
\end{eqnarray}
where 
\begin{eqnarray}
  s&=&\frac{1}{2}\{(1+\azetta{\braket{z_1}{z_2}})^{1/2}+
                 (1-\azetta{\braket{z_1}{z_2}})^{1/2}  \}, \\
  t&=&\frac{1}{2}\{(1+\azetta{\braket{z_1}{z_2}})^{1/2}-
                 (1-\azetta{\braket{z_1}{z_2}})^{1/2}  \}. 
\end{eqnarray}

\par \noindent 
From (\ref{eq:connection-form}) and the formulas above 
we can obtain the explicit form of $\cala$. We leave it to the readers. 

\par \noindent
Last let us give the explicit form to $\calf$ in 
(\ref{eq:local-curvature-form}) making use of (\ref{eq:matrix-form}) and 
(\ref{eq:v-dagger-v-inverse}) : 
\begin{eqnarray}
 \calf&=&\frac{1}{1-\szetta{\braket{z_1}{z_2}}}
 {\left(
    \begin{array}{cc}
      s& -\frac{\braket{z_1}{z_2}}{\azetta{\braket{z_1}{z_2}}}t \\
      -\frac{\braket{z_2}{z_1}}{\azetta{\braket{z_1}{z_2}}}t &s
   \end{array}
  \right)}
 {\left(
    \begin{array}{cc}
      F_{11}& F_{12}\\
      F_{21}& F_{22}
   \end{array}
  \right)} 
 {\left(
    \begin{array}{cc}
      s& -\frac{\braket{z_1}{z_2}}{\azetta{\braket{z_1}{z_2}}}t \\
      -\frac{\braket{z_2}{z_1}}{\azetta{\braket{z_1}{z_2}}}t &s
   \end{array}
  \right)}   \nonumber \\
 &\equiv& \frac{1}{1-\szetta{\braket{z_1}{z_2}}}
 {\left(
    \begin{array}{cc}
      {\calf}_{11}& {\calf}_{12}\\
      {\calf}_{21}& {\calf}_{22}
   \end{array}
  \right)} 
\end{eqnarray}
where
\begin{eqnarray}
{\calf}_{11}&=&\frac{1+\sqrt{1-\szetta{\braket{z_1}{z_2}}}}{2}
         \left\{1-\frac{\szetta{z_1-z_2}\szetta{\braket{z_1}{z_2}}}
         {1-\szetta{\braket{z_1}{z_2}}}\right\}d{\bar z_1}\wedge dz_1
           \nonumber \\    
       &-& \frac{\szetta{\braket{z_1}{z_2}}}{2}
            \left\{1-\frac{\szetta{z_1-z_2}}{1-\szetta{\braket{z_1}{z_2}}}
            \right\}d{\bar z_2}\wedge dz_1  \nonumber \\   
       &-& \frac{\szetta{\braket{z_1}{z_2}}}{2}
            \left\{1-\frac{\szetta{z_1-z_2}}{1-\szetta{\braket{z_1}{z_2}}}
            \right\}d{\bar z_1}\wedge dz_2   \nonumber \\
       &+& \frac{1-\sqrt{1-\szetta{\braket{z_1}{z_2}}}}{2}
         \left\{1-\frac{\szetta{z_1-z_2}\szetta{\braket{z_1}{z_2}}}
                  {1-\szetta{\braket{z_1}{z_2}}}
          \right\}d{\bar z_2}\wedge dz_2 \ ,  \nonumber \\
      &&{} \nonumber \\
{\calf}_{12}&=&-\frac{\braket{z_1}{z_2}}{2}
         \left\{1-\frac{\szetta{z_1-z_2}\szetta{\braket{z_1}{z_2}}}
         {1-\szetta{\braket{z_1}{z_2}}}\right\}d{\bar z_1}\wedge dz_1
           \nonumber \\    
   &+& \frac{\braket{z_1}{z_2}\{1-\sqrt{1-\szetta{\braket{z_1}{z_2}}}\}}{2}
            \left\{1-\frac{\szetta{z_1-z_2}}{1-\szetta{\braket{z_1}{z_2}}}
            \right\}d{\bar z_2}\wedge dz_1 \nonumber \\     
     &+& \frac{\braket{z_1}{z_2}\{1+\sqrt{1-\szetta{\braket{z_1}{z_2}}}\}}{2} 
            \left\{1-\frac{\szetta{z_1-z_2}}{1-\szetta{\braket{z_1}{z_2}}}
            \right\}d{\bar z_1}\wedge dz_2   \nonumber \\
     &-& \frac{\braket{z_1}{z_2}}{2}
         \left\{1-\frac{\szetta{z_1-z_2}\szetta{\braket{z_1}{z_2}}}
                  {1-\szetta{\braket{z_1}{z_2}}}
          \right\}d{\bar z_2}\wedge dz_2 \ ,  \nonumber \\
      &&{} \nonumber \\
{\calf}_{21}&=&-\frac{\braket{z_2}{z_1}}{2}
         \left\{1-\frac{\szetta{z_1-z_2}\szetta{\braket{z_1}{z_2}}}
         {1-\szetta{\braket{z_1}{z_2}}}\right\}d{\bar z_1}\wedge dz_1
           \nonumber \\    
     &+& \frac{\braket{z_2}{z_1}\{1+\sqrt{1-\szetta{\braket{z_1}{z_2}}}\}}{2}
            \left\{1-\frac{\szetta{z_1-z_2}}{1-\szetta{\braket{z_1}{z_2}}}
            \right\}d{\bar z_2}\wedge dz_1 \nonumber \\     
     &+& \frac{\braket{z_2}{z_1}\{1-\sqrt{1-\szetta{\braket{z_1}{z_2}}}\}}{2} 
            \left\{1-\frac{\szetta{z_1-z_2}}{1-\szetta{\braket{z_1}{z_2}}}
            \right\}d{\bar z_1}\wedge dz_2   \nonumber \\
     &-& \frac{\braket{z_2}{z_1}}{2}
         \left\{1-\frac{\szetta{z_1-z_2}\szetta{\braket{z_1}{z_2}}}
                  {1-\szetta{\braket{z_1}{z_2}}}
          \right\}d{\bar z_2}\wedge dz_2 \ ,  \nonumber \\
      &&{} \nonumber \\
{\calf}_{22}&=&\frac{1-\sqrt{1-\szetta{\braket{z_1}{z_2}}}}{2}
         \left\{1-\frac{\szetta{z_1-z_2}\szetta{\braket{z_1}{z_2}}}
         {1-\szetta{\braket{z_1}{z_2}}}\right\}d{\bar z_1}\wedge dz_1
           \nonumber \\    
       &-& \frac{\szetta{\braket{z_1}{z_2}}}{2}
            \left\{1-\frac{\szetta{z_1-z_2}}{1-\szetta{\braket{z_1}{z_2}}}
            \right\}d{\bar z_2}\wedge dz_1  \nonumber \\ 
       &-& \frac{\szetta{\braket{z_1}{z_2}}}{2}
            \left\{1-\frac{\szetta{z_1-z_2}}{1-\szetta{\braket{z_1}{z_2}}}
            \right\}d{\bar z_1}\wedge dz_2   \nonumber \\
       &+& \frac{1+\sqrt{1-\szetta{\braket{z_1}{z_2}}}}{2}
         \left\{1-\frac{\szetta{z_1-z_2}\szetta{\braket{z_1}{z_2}}}
                  {1-\szetta{\braket{z_1}{z_2}}}
          \right\}d{\bar z_2}\wedge dz_2 \ .
\end{eqnarray}
We have calculated the curvature form $\calf$. This one has an interesting 
structure (compare ${\calf}_{11}$ with ${\calf}_{22}$ and ${\calf}_{12}$ with
${\calf}_{21}$). The author believes strongly that our $\calf$ gives a 
solution of field equations of some non--commutative field theory.

\vspace{10mm}
\section{Discussion}

We have calculated Chern--characters for pull--back bundles on ${\cal D}_1$ 
and ${\cal D}_2$, and suggested a relation with some non--commutative field 
theory (or non--commutative differential geometry) for the case $m=2$. For 
the case $m=3$ we have left the calculations to the readers as a problem. 

\par \noindent 
Our paper is based on coherent states, but we believe that one can trace  
the same process for generalized coherent states (namely geometry of 
generalized coherent states).  See \cite{KF7} and \cite{KF8} for 
generalized coherent states. 
In the forthcoming paper \cite{KF9} we will report this. 

When the author was performing the calculations in section 3, the paper 
\cite{SV} appeared. It seems that our paper has some deep relation to it.

\vspace{1cm}
%
%%%%%%%%%%%%
%References%
%%%%%%%%%%%%

\end{document}